\documentclass [11pt,amsmath, amssymb, aps, bbm, nofootinbib, noindent, notitlepage] {article}
 \usepackage{lipsum}
\usepackage[margin=1.75cm]{geometry}
\usepackage{amssymb}
\usepackage{graphicx}
\usepackage{amsmath}
\usepackage[hidelinks]{hyperref}
\usepackage{array}
\usepackage{verbatim} 
\usepackage{multirow}
\usepackage{manfnt}
\usepackage{mathrsfs}
\usepackage{hyperref}
\usepackage{float}
\usepackage{caption}
\usepackage{braket}
\usepackage[sort, numbers]{natbib}
\usepackage{authblk}
 \usepackage{lipsum}

\usepackage{csquotes}

\numberwithin{equation}{section}

\usepackage{tocloft}

\usepackage{tocloft}

\def\be{\begin{equation}}
\def\ee{\end{equation}}

\usepackage{authblk}
\usepackage{color}

\title {Tightrope-walking rationality in action: \\ Feyerabendian insights for the foundations of quantum mechanics}
\author {Daniele Oriti}
	\affil{Arnold Sommerfeld Center for Theoretical Physics \\ Munich Center for Mathematical Philosophy \\ Ludwig-Maximilians-Universit\"at München, Germany \\ daniele.oriti@physik.lmu.de}

 \date{}

\begin{document}

\maketitle

\begin{abstract}
We scan Paul K. Feyerabend's work in philosophy of physics and of science more generally for insights that could be useful for the contemporary debate on the foundations of quantum mechanics. 
We take as our starting point what Feyerabend has actually written about quantum mechanics, but we extend our analysis to his general views on realism, objectivity, pluralism, and the relation between physics and philosophy, finding that these more general views could in fact offer many interesting insights for physicists and philosophers working on quantum foundations.
\end{abstract}

 \section{Introduction}
In the area of foundations of physics, the analysis of technical aspects of physical theories and experimental results has the purpose of elucidating the conceptual pillars of the theories, which we use, in turn, to interpret what we see in experiments. In some cases, work on these foundations impact only our understanding of theories. Thus, it is valuable but not ground-breaking; it is interesting for scholars, but without much broader implications. In other cases, it shakes the whole edifice of theoretical physics, and has important implications for fundamental issues in philosophy of science and philosophy more generally. Thus, it is interesting for a broader set of philosophers, alongside physicists. This was certainly the case, and it still is\footnote{It is also the case for the conceptual aspects of research on quantum gravity.}, for the foundations of quantum mechanics, whose puzzling aspects have provided exercises for prominent philosophers from its first developments. 

\
 
In the following, we focus on the contribution by P. K. Feyerabend (PKF) to the debate on the foundations and interpretation of quantum mechanics, on which he maintained a keen interest from early on but to which he contributed most actively in 1950s-1960s. The most relevant works are those collected in the first volume of PKF's papers \cite{PKF}, to which we will refer throughout (rather than referring to the original publications). 

We will discuss several interesting points made by PFK. We group them in three categories. The first is about the general relation between physics and philosophy, which also serves as a framework for the philosophical analysis of quantum mechanics (and philosophy of physics at large). The second concerns the debate between realism and instrumentalism, applied to the interpretation of quantum mechanics, but also, again, as two broader perspectives on scientific theories. The third deals with specific insights  on quantum mechanics itself, concerning the interpretation of quantum states, probabilities, Bohr's view, and hidden variables.  

A result of our scan, beyond specific insights on quantum mechanics, is to illustrate how PFK constantly aimed to achieve a clearer understanding of the various issues he was facing, finding compelling rational arguments in favor of one particular viewpoint, while at the same time avoiding any form of dogmatism, even regarding the rational arguments he had just put forward, and regarding rationality itself (or the criteria adopted for evaluating his own arguments). This struggle characterizes his own version of rational analysis, dubbed \lq tightrope-walking rationality\rq ~by Farrell \cite{farrell}, which we see in the context of the foundations of quantum mechanics and related philosophical issues. The dogmatic version of the two sides of any argument, e.g. realism vs instrumentalism, is what we see on the right and on the left of the tightrope we are forced to walk on, and we should avoid falling toward either side. The need to avoid dogmatism of any sort appears to be the strongest driver in PKF's analysis, to the point where any perspective that is at risk of becoming a dogmatic position or facilitating a dogmatic attitude is immediately objected to {\it exactly on that basis}. Thus, avoiding dogmatism turns quickly from a normative methodological rule for philosophical inquiry into an evaluative criterion of philosophical positions in itself. We will emphasize various concrete instances of application of such tightrope-walking rationality in what follows.

\

Before we start, it is worth emphasizing what this contribution is about and what is not aiming at. It is not intended as a scholarly contribution about PKF's work and thought. We do not attempt a precise or detailed analysis of what  PKF really meant or of how PKF's thinking evolved over the years. We do not frame PKF's thinking within the historical debate on quantum foundations, and we do not assess his actual impact on the same debate. For proper scholarly work on PKF, especially in relation to contemporary debates in philosophy of science (e.g. on realism, perspectivism, pluralism,  scientific methodology, etc), see for example \cite{preston, modern1, modern2} or the cited \cite{farrell}, and, in broader terms \cite{Oberheim2006,Tsou}. For further analysis on PKF's philosophy of physics, see instead \cite{vanStrien,Kuby2021,manchak,Kuby2016,Kuby2022,DelSanto}.
What we aim to do is much more limited. Being interested ourselves in the many puzzles of quantum foundations, we scan PKF's writings for useful insights that could help us improving our own understanding of them, and could suggest ways to make technical or conceptual progress. In particular, we ask PKF for advice about how one should seek some form of scientific realism behind quantum mechanics, about the complex relation between a scientific theory and reality, and about how to be critical toward our own attitudes when presented with a new set of phenomena and a new candidate description of them. Also in this respect, our contribution should be seen as a modest addition to the scholarly work on this topic, examples being \cite{vanStrien,Kent}
It will be, then, an exercise in picking up useful inputs for current research, and suggestive stimuli to be put to fruition at a second stage. We will try, of course, to be careful in our reading and understanding PKF's thoughts, and to avoid misinterpretations or cherry-picking or other forms of unfaithfulness to PKF's intentions and ideas. In case we fail despite our best efforts, and some cherry-picking turns out to be happening, even if unintended,  we hope that at least the cherries are good.

 \section{Relation between physics and philosophy, in the foundations of QM}
Work on the foundations of physics involves tackling also problems in the philosophy of physics, and philosophy of science more generally. This is especially true in the context of the foundations of quantum mechanics. Quantum mechanics indeed challenges fundamental pillars of our modeling and understanding of the physical world, that percolate into our basic metaphysics. The central role that quantum mechanics plays in science, moreover, makes it impossible to neglect such challenges. The result is that the foundations of quantum mechanics are the shared playground of both physicists and philosophers, each with their own set of tools and objectives. While both physicists and philosophers can in principle hide behind the technicalities of their disciplines, the way they approach each specific issue is inevitably influenced by how they see the general relation between physics and philosophy. PKF has some insights to offer on their dialogue and joint work, which is useful also to frame better his own work in this domain.

\

An immediate point is PKF's preference for philosophical analyses that, when not directly done by physicists, it is proceeding in parallel with scientific developments and to some extent guided by them; this point has been emphasized already in \cite{Kuby2021,Shaw2021}. On the one hand, an important goal of the philosophy of physics is, in PKF's view, to contribute to the scientific progress; on the other hand, a physical argument or result that supports or suggests a certain philosophical perspective becomes the best argument in its favor and has more weight than abstract philosophical reasoning going in the same or opposite direction. This applies to metaphysical issues, as well as epistemological or methodological ones.
In this sense, one can understand the many occasions in which PKF expresses discomfort or outright annoyance with philosophical positions that appear oblivious to the hard-won results by scientists or that are based on a superficial picture of scientific theories, and of how they have been obtained and tested, or with methodological rules that  have no relation with the practice of science.

We would suggest that a way to understand the spirit of PKF's attitude in this regard is to consider his tightrope-walking rationality not so much as an abstract theory of rationality to be applied (or imposed) onto science and scientists, but, quite the contrary, as a model of rationality deduced or abstracted from the way actual scientists work, from the necessarily fallibilist and wavering proceeding of real-world science.  

In a general survey on scientific realism in physics, for example, PKF praises the ideas of Maxwell and Mach because \lq\lq They are also more subtle. They were developed in close connection with research [...]. But one feels a sense of relief when transferred from the fruitless technicalities and ontological primitivisms of modern 'philosophers' to the brief, simple, but profound remarks of these scientists\rq\rq. [\cite{PKF}, page 11].
 
And here is PKF in one of his many remarks about the role of scientific methodology in relation to science, in the context of his reappraisal of Bohr's view:
 
 \begin{quote}
Lakatos says I use \lq some inconsistencies and waverings in Bohr's position for a crude apologetic falsification of Bohr's philosophy\rq. This is the language of the rationalist who looks at research from the safe distance of his office and wants to run science according to one method and base it on one philosophy and who fails to understand that researchers adapt their ideas to the always changing research situation. Bohr does not \lq waver\rq because he does not know what to do, but because new discoveries are constantly being made and because he takes them into account. - \cite{PKF}, page 294, footnote 100
 \end{quote}
 
 One cannot but relate the \lq wavering\rq ~of the scientist who is trying to make sense of the world in light of new evidence and the tightrope-walking of a philosophers who needs to avoid dogmatisms of all kinds.
 
 Ultimately, physical arguments (and results) trump philosophical prejudices, be they methodological or ontological. This is as it should be, to make science progress. On the methodological side, this attitude is also at the basis of PKF's methodological anarchism \cite{AgainstMethod}, since it allows for different methodological prescriptions to be employed in different circumstances, and any given methodological rule to be violated, if deemed useful to achieve scientific progress (a similar understanding of PKF's methodological anarchism has been advocated in \cite{Shaw2017}). 

Does this mean that philosophy plays no role in scientific progress? Obviously not. Philosophical arguments can guide or support physics; a certain philosophical perspective can prove fertile in driving scientific progress or solving conceptual puzzles encountered by scientists. Ontological preferences are means to physical understanding; they are either starting points for new scientific constructions or results of (analysis of) existing ones. Philosophical arguments can also be a good motivation to oppose a scientific theory or viewpoint, but only as starting point to develop alternative scientific theories. It is at the level of physics that the debate will have to be settled, with the physical reasons being given appropriate (and greater) weight. Philosophical arguments can win a philosophical battle, never one about physics. This implies, in particular, that philosophical as well as physical arguments should be considered in relation to specific theories, those that we are scrutinizing for viability, and not in general, abstract terms.   
 
 PKF is very clear on this point, when discussing the debate around the heliocentric view and then using it to elucidate the parallel debate on the realist vs instrumentalist interpretation of quantum theory (see for example \cite{PKF}, page 185-186). In \cite{Shaw2020}, PKF's position regarding the \lq realism vs instrumentalism\rq debate in discussed from a similar understanding of his attitude.
 

This is a warning that should be raised also in front of physicists trying to wrap their theories in fancy philosophical clothes, trying to make their insights look deeper, or trying to call philosophical frameworks in further support of their physical theories. This move can have value but one has to be very careful.  Bad philosophical arguments can end up obscuring good physical motivations and bring unnecessary trouble to a physical theory or viewpoint that could have stood on more solid grounds without. This is what has happened, according to PKF, to the "instrumentalist" ~interpretation of quantum mechanics. In the attempt to further justify such position beyond the physical arguments by linking it with more general positivism, quantum physicists ended up offering easy points of attack to their opponents. Attacking their positivist position and refuting it with good philosophical arguments made their view of quantum theory appear weaker, despite their being in possession of strong physical results supporting it. So strong, in fact, that they did not need to be further justified in terms of any positivist philosophy. In the same spirit, PKF warns about expressing or interpreting physical theories in a language adapted to or coming from a specific philosophical perspective. In the specific case of quantum transitions, for example, PKF insists that the absence of intermediate states is something to be argued for because of the empirical evidence and of the content of quantum theory that successfully accounts for it, while arguing for this conclusion or justifying the theory on the basis of a philosophical doctrine (here, positivism) does not help: 
 
 \begin{quote}
[...] 
the quantum postulate does not merely exclude the knowledge of, or the observability of, the intermediate states; it excludes these intermediate states themselves. Nor must the argument be read as asserting, as is implied in many presentations by positivist-minded physicists, that the intermediate states do not exist because they cannot be observed. For it refers to a postulate (the quantum postulate) which deals with existence, and not with observability. [...] Physicists who have adopted the positivistic principle that things which cannot be observed do not exist try to justify the indefiniteness of state descriptions by a combined reference to the fact that they cannot be observed and to this principle. Philosophers immediately expose the fallacy of the argument (if they are anti-positivists, that is) and think that they have thereby shown the existence, or at least the physical possibility, of sharp states. This is, of course, not correct [...]
\cite{PKF}, page 187-190
\end{quote}

On the contrary, the successful theory can be used to support a given philosophical position, here instrumentalism, as PKF notes after comparing again the situation in QM with the one concerning the heliocentric theory and Bellarmine's objections to it:
\begin{quote}
[...]  modern physics has found new physical reasons why its own most important theory, the quantum theory, cannot be anything but an instrument of prediction. These reasons are of precisely the same character as were Ptolemy's: a realistic interpretation of the quantum theory is bound to lead to incorrect predictions. Admittedly, in the arguments usually presented these physical difficulties are almost buried beneath an unacceptable positivism. However, this does not mean that they do not exist [...] 
- \cite{PKF}, page 187-190
\end{quote}

 A purely philosophical critique of an instrumentalist interpretation of QM and, more generally, of any given interpretation of a physical theory motivated by physical arguments or observations, and, more generally still, of any physical theory that seems to contradict our philosophical convictions, is sterile and irrelevant. In the specific case, a realist interpretation of QM can only be taken seriously if physical arguments in its support can be found or if it is used to construct a better theory. The latter, more difficult task would constitute real scientific progress and it is what opponents of a given viewpoint (or theory) are called to embark on.
 
 \begin{quote}
 the instrumentalism in the quantum theory is not a purely philosophical affair that can be disputed away by general arguments in favour of realism. 
 A realistic alternative to the idea of complementarity is likely to be successful only if it implies that certain experimental results are not strictly valid. It therefore demands the construction of a new theory, as well as demonstration that this new theory is experimentally at least as valuable as the theory that is being used at present. This is a formidable task indeed, and a task that is not even recognized by the purely philosophical champions of realism in microphysics.
- \cite{PKF}, page 193-194
\end{quote}

In fact, seriousness accepting the task of constructing a better theory than QM, on the basis of a more convincing philosophical perspective, and a theory that could bring new physical insights and new predictions, is what PKF clearly appreciated in Bohm's work on a new theory based on "hidden variables" and the pilot wave (following de Broglie). In accordance with his views on the respective role of physics and philosophy in the development and interpretation of physical theories, PKF maintained this appreciation {\it despite} his strong criticisms of the philosophical arguments and ideas of Bohm himself:    

\begin{quote}
Bohm's physical ideas are original, refreshing, and sorely needed in a time of complacency with respect to fundamentals. But the philosophical standpoint taken up with respect to both physics and cosmology is traditional, and perhaps even reactionary - \cite{PKF}, page 220
\end{quote}
 
In the same discussion, PKF introduces another important element: a sound position can quickly become counterproductive and represent an obstacle to further progress the very moment it becomes dogmatic, no matter how many solid arguments support it. A position that is well supported by physical arguments and observations, that is embodied in the best physical framework we were able to find, and that it is proving fertile for further scientific progress  becomes dogmatic when it is declared to be necessarily part of any future theory, to stand without {\it possible} alternatives, rather than merely current ones, as something having absolute validity. Also here one can see the ideal tightrope-walking philosopher or scientist moving the balance quickly away from the full (and possibly uncritical) embracing of a theory and towards a more skeptical or simply actively critical attitude, to avoid falling into the pit of dogmatism. This dogmatic turn is what PKF attributes to the "orthodox" Copenhagen interpretation and to the physicists holding it. In his words: 

\begin{quote}
The idea of complementarity can be interpreted [...]
as an attempt to provide an intuitive picture for an existing theory, viz. wave mechanics, and as a heuristic principle guiding future research. This interpretation is undogmatic as it admits the possibility of alternatives, and even of preferable alternatives. A physicist who looks at complementarity in this way will regard it as an interesting fact about quantum theory that it is compatible with a relational point of view where interaction is a necessary condition of the meaningful applicability of terms which within classical physics (relativity included) are definable without such reference. He will also point out that there exist no satisfactory alternatives. But he will never go as far as to assert that such alternatives will never be found, or that they would be logically inconsistent, or that they would contradict the facts. But Bohr's idea of complementarity can also be interpreted [...]
as a basic philosophical principle which is incapable of refutation and to which any future theory must conform. Bohr himself most certainly took this stronger point of view. - \cite{PKF}, page 221-222
\end{quote}

and

\begin{quote}
it has been attempted, both by Bohr, and by some other members of the Copenhagen circle, to give greater credibility to these ideas by incorporating them into a whole philosophical (ontological) system that comprises physics, biology, psychology, sociology and perhaps even ethics. Now the attempt to relate physical ideas to a more general background and the correlated attempt to make them intuitively plausible is by no means to be underestimated. Quite the contrary, it is to be welcomed [...] 
However, the philosophical backing of physical ideas that emerged from these more general investigations has led to a situation that is by no means desirable. It has led to the belief in the uniqueness and the absolute validity of both of Bohr's assumptions [...]
viz. the indefiniteness of state descriptions and the relational character of the quantum-mechanical states, which, so it is added, cannot be replaced by different ideas without creating formal inconsistencies, or inconsistencies with experiment. - \cite{PKF}, page 313
 \end{quote}
 
We will further discuss some of these statements in relation to specific insights of PKF regarding Bohr's views. Here, we conclude this general discussion with three remarks. 
First, to better appreciate PKF's position above, it is useful to recall that PKF is writing in a historical period in which the debate about quantum foundations and interpretations had almost vanished, or at least it was not prominent among scientists anymore; a time in which the orthodox \lq\lq Copenhagen\rq\rq interpretation, an instrumentalist or maybe simply a dismissive attitude toward any philosophical debate about physical theories, was entirely dominant in the physics community.   
Second, the problematic link between the risk of epistemological and scientific dogmatism and the use of \lq\lq no alternative arguments\rq\rq in theory assessment and, more generally, the delicate interplay between scientific development and philosophy of physics, have recently been discussed in the context of the debate on non-empirical theory assessment in fundamental physics \cite{NoAlternative, WhyTrust, NoAlternativeProliferation}.
Finally, while PKF criticizes Bohr as partially responsible for the dogmatic turn of the Copenhagen position, he was more generally full of praise, especially in later writings, for Bohr's methodological and epistemological attitude, as providing a lesson about the proper interplay between physics and philosophy and for the need to avoid dogmatism.

Indeed, contrasting Popper with Bohr, PKF writes:

\begin{quote}
The idea of complementarity is therefore not just the result of having pursued a mistaken programme to the bitter end as Popper would want us to believe. Bohr, after all, did consider a purely statistical theory. He did consider such a theory despite the fact that it was not in line with his own point of view [...] 
It was only after the refutation of this theory that he returned to his earlier philosophy - and this time with very good reasons. - \cite{PKF}, page 268
\end{quote}

and, on Bohr's attitude toward quantum physics itself and the role of philosophical reasoning in understanding it (questioning also his holding any naive "instumentalist" position): 

\begin{quote}
(quoting Heisenberg) 'Bohr was primarily a philosopher, not a physicist, but he understood that natural philosophy . . . carries weight only if its every detail can be subjected to the . . . test of experiment'.
His approach differed from that of the school-philosophers whom he regarded with a somewhat 'sceptical attitude, to say the least' and whose lack of interest in 'the important viewpoint which had emerged during the development of atomic physics' he noticed with regret. But it also differed, and to a considerable degree, from the spirit of what T. S. Kuhn has called a 'normal science'. Looking at Bohr's method of research we see that technical problems, however remote, are always related to a philosophical point of view; they are never treated as 'tiny puzzles' whose solution is valuable in itself, even if one has not the faintest idea what it means, and where it leads. [...] Emphasis is put on matters of principle and minor discrepancies, or 'puzzles' in the sense of Kuhn, instead of being deemphasized, and assimilated to the- older paradigm, are turned into fundamental difficulties by looking at them from a new direction, and by testing their background 'in its furthest consequences by exaggeration'. 
- \cite{PKF}, page 269-274
\end{quote}

and again:

\begin{quote}
[...] throughout the period of the older quantum theory Bohr underlined the 'formal' character of the new ideas which arose in connection with the principle of correspondence: these ideas, we can read repeatedly in his earlier papers, lead to some surprising predictive successes, and they also establish some order in the steadily accumulating empirical material. However, they do not form a theory in the ordinary, or 'classical' sense of the word, that is, they do not contain a coherent description of new and objective features. Nor can we say that they explain the relevant experimental facts. All this is emphasized by Bohr himself and not only once, but in almost every paper he writes. He also emphasizes that the dependence on classical ideas that is implied by the use of the correspondence principle characterizes 'the present state of science', or 'the present state of our knowledge', and not the essence of physics. It is even regarded as somewhat undesirable. In these early years Bohr's aim seems to have been the gradual construction of an effective instrument of prediction which he hoped would act as a stimulus and as an aid towards the invention of a new atomic theory in the full sense of the word. [...] 
- \cite{PKF}, page 279-280
\end{quote}

Once more, PKF is arguing against a way of doing philosophy of physics which does not properly take into account the physical arguments leading to a theory or an interpretation of it, and it is thus based on an oversimplified picture of how science works. He does so by stressing how Bohr arrived at his "instrumentalist" interpretation after a long, undogmatic and serious struggle with physical, conceptual and observational results, and after having seriously considered possible alternatives to the same interpretation.\footnote{In fact we may note that Bohr's complementarity, and the associated contextual understanding of probabilities in QM, could be argued for on the basis of an analysis of the nature of probabilities and relative frequencies in the line of Boole's \lq conditions of possible experience\rq ~\cite{boole}, and thus again not a purely philosophical position, against Popper (but also the earlier Feyerabend).}

 \section{Realism vs instrumentalism, in the interpretation of QM}
The philosophical issue at the core of PKF's reflections in the context of the foundations of QM, and in fact of much of his (early) production, is the debate between realism and instrumentalism. This ties directly with the issue of how physical theories should be understood and used, of their descriptive or representational role, in addition to their use as tools for prediction of future experiences. We will not try to give a comprehensive summary of PKF's position. We will only briefly single out a couple of conclusions from his reflection, that we find useful for the foundations of QM and only discuss them in this context.

Roughly, (scientific) realism is the view that the world and its constitutive objects, together with their properties, exist independently of our knowledge of them, and scientific theories  somehow capture this independent reality. In this sense, scientific theories tell us 'what is real' out there, whether we observe or know about it or not.  Different versions of realism would then be identified in terms of how they articulate the correspondence between scientific theories and the world, and what they take this correspondence to imply about the theories. Establishing this correspondence requires relating the content of scientific theories to the results of our observations, our experiments, our experiences of the world, i.e. what can be labeled \lq the phenomena\rq. However, the correspondence will include also elements of the theories that have only indirect relation to phenomena. On the other hand, instrumentalism  takes scientific theories to be pure instruments for prediction and organization of phenomena (i.e. again, the result of direct observations), and judged only in terms of this. Thus, it denies any relevance to elements of the theories which are not direct relation to such phenomena and, more generally, any meaningful sense in which scientific theories \lq tell us what is real\rq in the world. They are tools, they do not represent anything beyond what we experience \footnote{Of course, here we are simplifying, and somewhat \lq straw-manning\rq, both realism and instrumentalism as philosophical positions; both can be much more nuanced and complex, as well as more mutually conciliatory, in the actual formulation and arguments of the scientists and philosophers who argue for such positions. This warning applies also to some of the considerations of PKF, and ours, in what follows.}. 

In the QM case as discussed by PKF, the debate between realism and instrumentalism is directly translated into the debate between the \lq Copenhagen\rq and the \lq hidden variables\rq interpretation of the QM formalism. This reflects the historical moment in which PKF was writing and the landscape of QM interpretations of the time. A certain standardized version of the QM interpretation coming from the Copenhagen school of Bohr, Heisenberg etc had become orthodoxy, and simplified from the richer variety and conceptual subtleties of the initial positions to become indeed a rather naive, and dogmatic, instrumentalist attitude. This reflected also a general dismissal of conceptual and foundational work on physical theories that, together with increased specialization and separation of physics and philosophy, became dominant in the post-war period (for an historical analysis and some examples, see \cite{postwar}). Similarly, realism in the context of QM is understood in a rather restricted sense. At the moment of PKF writings, the only well-developed alternative to the orthodox view was the deBroglie-Bohm mechanics\footnote{For a survey of interpretations of QM, see \cite{cabello}, while for their individual tenets and history, see the references cited therein and the corresponding entries in the Stanford Encyclopedia of Philosophy.} based on position variables for particles understood as in classical Newtonian mechanics (but supplemented with the new "pilot-wave" suggested by QM), and indeed the "objective reality" threatened by QM and defended by "realists" was the traditional one described by Newtonian mechanics. We will say more about modern perspectives on QM in the next section, but note here that there are more ways, nowadays, to be a realist in the context of QM (including both new proposed  ontologies, based for example on ontic quantum states, like the many-worlds interpretation \cite{MWI} and new ways of understanding realism itself \cite{realism1,realism2}), and that the Copenhagen perspective has evolved to be much richer and subtle than any straightforward instrumentalism (which can still be found defended with good arguments, though \cite{Peres}). 

As said, there are several versions of scientific realism. PKF presents (some of) them in the introduction to \cite{PKF}, showing how the nature of scientific theories and their relation to "facts" is in fact subtle, since theories are more complex than any naive picture of them as "candidate portraits of reality" but also as "systematization of observations" would want. This complexity should be properly appreciated  to tackle the issue of realism vs instrumentalism in a fruitful manner, and this is indeed the preliminary goal of PKF's analysis. In doing so, characteristically, PKF builds on the work of philosophers but even more on the practice of science and the insights of scientists. His remarks about the brilliance of Maxwell's and Mach's thinking were written in this context. Again, what is particularly appreciated is the pragmatic, progress-oriented attitude of the scientists, opposed to the dogmatic, principles-first attitude of (some) philosophers, but also of scientists heavily influenced (in PKF's view) by positivist philosophy. 

Here are PKF's words:

\begin{quote}
The first version of scientific realism therefore does not lead to a realistic interpretation for all theories, but only for those which have been chosen as a basis for research. It may be asserted (a) that the chosen theory has been shown to be true or (b) that it is possible to assume its truth, even though (ba) the theory has not been established or (bb) is in conflict with facts and established views.[...] A second version of scientific realism assumes that scientific theories introduce new entities with new properties and new causal effects. [...]  a direct application of the second version of scientific realism ('theories always introduce new entities') and a corresponding abstract criticism of' positivistic' tendencies are too crude to fit scientific practice. - \cite{PKF}, page 5-7
\end{quote}

\begin{quote}
Naive realists - and many scientists and philosophers supporting the second version belong to this group - assume that there are certain objects in the world and that some theories have managed to represent them correctly. These theories speak about reality. 
The accounts just given assume two different domains, or layers. On the one side we have phenomena, facts, things, qualities as well as concepts for the direct expression of their properties and relations. On the other side we have an abstract (quantitative) language in which the 'phantom pictures,' i.e. scientific theories, are formulated. The pictures are correlated to the phenomena, facts, things, qualities of the first domain. Attention is paid to the language of the pictures or the 'theoretical language', [...] and one considers ways of modifying and improving it. Little attention is paid to the 'observation language'. - \cite{PKF},  page 8-11
\end{quote}

PKF is emphasizing that scientific theories have many components, with different roles and different, subtle relations to (empirical) facts, constructed via a mixture of a priori assumptions and hypotheses and of empirical feedback from observations, and a highly non-trivial intertwining of conceptual ingredients, mathematical language and observational inputs.

\begin{quote}
A physical hypothesis does provide a guide and it also keeps the subject matter before our eyes. However, it makes us see the phenomena 'only through a medium'. Maxwell seems to fear that physical hypotheses may be imposed upon the phenomena without the possibility of checking them independently. As a result we cannot decide whether the phenomena are correctly represented by these hypotheses. Analogies avoid the drawbacks of mathematical formulae and of physical hypotheses. They are hypotheses in Mill's sense of the word, i.e. assumptions about the nature of things which have been examined and have passed tests. They have heuristic potential, but they don't blind us. 
- \cite{PKF}, page 11-12
\end{quote}

Moreover, it is a mistake, PKF argues, to believe that theories simply \lq\lq represent\rq\rq ~the world. This representational feature is questionable both for the directly observational components, which provide the basis for an instrumentalist interpretation, and for the more abstract components that we may take as indicating a \lq\lq hidden\rq\rq (and maybe truer) physical reality, from a realist standpoint.
Which aspects or parts of scientific theories do in fact represent is a choice grounded in physical understanding, not an immediately obvious one. One could say that it is {\it concepts} that represent, mediated, in case, by (mathematical) theories. This is a point that PKF makes also referring to Bohr's position (see also \cite{Curiel}).

Having appreciated such richness, one starts doubting the key distinction on which the realism vs instrumentalism debate is built, i.e. the \lq\lq double layer" constituted by "theories" on the one hand and "facts" on the other, as well as about the associated distinction between \lq\lq theoretical language" and \lq\lq observational language", i.e. the \lq\lq double language\rq\rq model. This is the first target of PKF's philosophical criticism. It becomes, in particular, a criticism of the instrumentalist position, which relies crucially on the above distinction in its insistence that only observations and empirical data should matter in our development and acceptance of physical theories. This is a part of PKF's broader criticism toward empiricism and especially the positivist tradition (for more details beyond the above sketch, see again the introduction of part 1 in \cite{PKF}).
 
PKF argues that all concepts, including those used to formulate facts, are theoretical concepts. It is theories that influence what we see as 'facts' and how we understand/interpret them. This is the well-known point of theory-ladenness of observations \cite{theory-laden}, by now accepted in philosophy of science (thanks to PKF's, Kuhn's and Hanson's work, plus the earlier work of Popper and the later work of Laudan, to mention just a few). It follows that any sharp or absolute distinction between observations and theories, between empirical facts and theoretical constructions, is false and misleading. Thus, a pure instrumentalist position is either disingenuous or, in some sense, a realist position in disguise, in which one accepts some (elements of some) theories as pointing to something real and others as tools only. It is also, to some extent, an overly conservative position, because it ends up retaining, without sufficient questioning, what are ultimately the conceptual theoretical elements of older scientific theories or philosophical traditions. 
 
 \begin{quote}
 According to Mach it is our task not only to classify, correlate and predict phenomena, but also to examine and to analyse them. And this task is not a matter for philosophy, but for science. [...] Science explores all aspects of knowledge, 'phenomena' as well as theories, 'foundations' as well as standards; it is an autonomous enterprise not dependent on principles taken from other fields. This idea according to which all concepts are theoretical concepts, at least in principle, is definitely in conflict with the positivistic version of scientific realism and it is very close to the point of view of ch. 2.6, thesis I.. - \cite{PKF}, page 13
\end{quote}

\begin{quote} 
The double language model 'clarifies' the distinction by cutting it off from scientific research and reformulating it in epistemological, i.e. non- scientific, terms. A 'clarification' is certainly achieved - simpleminded notions are always more easy to understand than complex ones — but the result has little to do with scientific practice. This is my main criticism of the double language model.
To elaborate: there is no doubt that the double layer model which scientists discussed in the last century captured certain features of scientific knowledge. The concepts used on the observational level are often quite different from the 'theoretical entities' of a newly introduced abstract theory - after all, they belong to an earlier stage of knowledge, they are familiar, their application may be connected with perceptual processes while the application of theoretical terms, especially of newly introduced theoretical terms, is mostly perception free (cf. the explanations in ch. 2.1). But a closer looks reveals that the situation is much more complex. - \cite{PKF}, page 13-14
 \end{quote}
 
The conclusion of PKF's reasoning is that the observational language (which is the basis for empiricism, then instrumentalism, but also certain realist positions) is also a theoretical one and should be treated as such.

\begin{quote}
[...] we may now tentatively put forward our {\it thesis I: the interpretation of an observation language is determined by the theories which we use to explain what we observe, and it changes as soon as those theories change}. [...] 
(1) According to thesis I, we must distinguish between appearances (i.e. phenomena) and the things appearing (the things referred to by the observational sentences in a certain interpretation). This distinction is characteristic of realism.
(2) The distinction between observational terms and theoretical terms is
a pragmatic (psychological) distinction which has nothing to do with the
logical status of the two kinds of term. On the contrary, thesis I implies that
the terms of a theory and the terms of an observational language used for
the tests of that theory give rise to exactly the same logical (ontological)
problems. 
[...]
- \cite{PKF}, page 31-33
\end{quote}

Not even everyday language should be taken for granted, accepted as indispensable element of our epistemology, and of our science, and taken as the basis for accepting a certain observational language. The reasons are that, first of all, there is no \lq\lq uniform everyday language\rq\rq ~but a mixture of languages \lq\lq  which has received its interpretation from various and often incompatible and obsolete theories\rq\rq. Second, \lq\lq terms which at some time were regarded as observational elements of 'everyday language' (such as the term 'devil') are no longer regarded as such. Other terms, such as 'potential', 'velocity', etc., have been included in the observational part of everyday language, and many terms have assumed a new use.\rq\rq, i.e. what is part of everyday language changes with the adoption of new theories (quotes from \cite{PKF}, page 31).

This criticism  of the double-layer and double-language model has immediate consequences for our understanding of QM, and for the debate between \lq\lq Copenhagen" and \lq\lq hidden variables". 
 
 The criticism is directed in particular at the conviction (central in the Copenhagen position) that our observations have to be expressed in classical terms, i.e., those that are, upon scrutiny, inherited from classical Newtonian mechanics. That is, not only observations are theoretically loaded, as we agreed, but they are loaded by an older theory. There may be good reasons to accept this (implicit) theoretical basis, but they have to be argued for, not taken for granted. We note that this reliance on classical mechanics in providing the necessary observational language for QM is a key ingredient not only of the traditional Bohr/Copenhagen position, but also of its more modern incarnation \cite{AuffevesGrangier}. According to PKF, this is a remnant of a hard-to-shake positivist influence. 
 The danger of such uncritical acceptance is not only that it unnecessarily restricts our epistemological and scientific options, and dampens our creativity, but that it may quickly turn dogmatic. On the contrary, we should always leave open (and in fact actively pursue) the possibility of revising our theoretical structures, including those that constitute the basis of our observational language and those that follow from older theoretical schemes. PKF is very forceful in attacking this aspect of the Copenhagen orthodoxy, maybe insisting on a not-too-charitable interpretation of Bohr's own words (which could be taken to be simply stressing the new lessons of QM and the need to take them seriously in future developments, not necessarily maintaining them unaltered):

\begin{quote}
 Bohr seems to assume that this will hold for any future theory of microscopic entities. Now it may be conceded that the laws of quantum mechanics do not admit of a straightforward interpretation on the basis of a classical model, as such a model would be incompatible either with the principle of superposition or with the individuality of the microscopic entities. It may also be conceded that as a matter of fact we do find it difficult (though by no means impossible) to form an intuitive picture of processes which are not dependent upon the classical framework. But from this psychological predicament we can by no means infer (assumption 1) that such intuitive understanding will never be possible. And it would be even less correct to assume on that basis that the concept of a non-classical process cannot be formed (assumption 2); for it is well known that we can form and handle concepts even of those things which we cannot readily visualize.  [...] according to Bohr it would even be a 'misconception to believe that the difficulties of the atomic theory may be evaded by eventually replacing the concepts of classical physics by new conceptual forms', as there exist 'general limits of man's capacity to create concepts'. - \cite{PKF}, page 23-24
 \end{quote}

The Copenhagen position, thus, according to PKF, gives a good example of how philosophical attitudes can both facilitate scientific progress and hamper it especially when they leave part of our epistemic tools (here, the observational language) unquestioned . Indeed, it follows directly from specific philosophical ideas: 
\begin{quote} 
The first idea is that the belief in classical physics has influenced not only our thinking but also our experimental procedures and even our 'forms of perception'. This idea gives a correct description of the effect which the continued use of a fairly general physical theory may have upon our practices and upon our perceptions: it will become increasingly difficult to imagine an alternative account of the facts. The second idea is inductivism. According to inductivism we invent only such theories as are suggested by our observations. Combined with the first idea inductivism implies that it is psychologically impossible to create non-classical concepts and to invent a non-classical 'conceptual scheme'. The third idea is the principle of pragmatic meaning. According to this idea, the use of classical methods and the existence of classical 'forms of perception' imply that the observation language possesses a classical interpretation (see above). As a non-classical picture of the world would lead to an interpretation which is inconsistent with this classical interpretation, such a non-classical picture, apart from being psychologically impossible, would even involve a logical absurdity. [...] As opposed to this it is sufficient to point out that even in a situation where all facts seem to suggest a theory which cannot any longer be maintained to be universally true, that even in such a situation the invention of new conceptual schemes need not be psychologically impossible so long as there exist abstract pictures of the world (metaphysical or otherwise) which may be turned into alternative interpretations.  - \cite{PKF}, page 23-24
\end{quote}

PKF's target is the orthodoxy of the time, and its instrumentalist and positivist core. We note here, though, that we could direct a similar criticism, and with the same philosophical arguments, toward any realist interpretation of QM in which the physical reality we decide to preserve, despite the challenges of QM, corresponds to an ontology built on the same premises of our observational language or our common sense, that is, upon reflection, on classical Newtonian mechanics. An example would be the same deBroglie-Bohm theory that was the only contender of the orthodox Copenhagen view at the time of PKF's writing. The criticism would then extend to more modern perspectives, Bohmian or not, aiming at re-establishing a primitive ontology of particles moving in space and in presence of an absolute time, an ontology which is essentially Newtonian \cite{PrimitiveOntology}, at the heart of fundamental physics. Relatedly, it is interesting to look at modern "hidden variables" completions of QM, formalized in the mathematical framework of "ontological models" \cite{Leifer}, where quantum states are epistemic states (states of knowledge) over a hidden (from QM) set of ontic states of the physical system under consideration. These models (which at present can cover only "fragments" of QM)  are indeed forced by the lessons of QM (including both theoretical developments of the last decades and its observational successes) to look for a precise characterization of the ontic structure of the world (what is real) in terms much farther away from classical realism and a Newtonian picture than the early advocates of a realist alternative to the Copenhagen instrumentalism (and the Bohmians) would have liked (see, e.g. \cite{spekkens}).
 
It is instructive, then, to read further PKF's critical position about this dogmatic turn of empiricism, leading to the instrumentalist Copenhagen interpretation of QM, appreciating the anti-dogmatic spirit, but also putting the realist Bohmian or hidden variable position under the same critical lens (we could try to substitute any reference to Bohr or Copenhagen in this quote with a reference to Einstein or Bohm or other "realist" proposal), to realize that the dogmatism attacked by PKF has spread beyond the Copenhagen interpretation.

The historical example that PKF uses for support of his statement that experimental evidence can and should be questioned, as it implicitly encodes older theories, is his favourite Aristotelian vs Galilean clash:

\begin{quote}
The Galilean tradition, as we may call it, therefore proceeds from the very reasonable point of view that our ideas as well as our experiences (complicated experimental results included) may be erroneous, and that the latter give us at most an approximate account of what is going on in reality. Hence, within this tradition the condition to be satisfied by a future theory of the microcosm is not that it be simply compatible with duality and the other laws used in the above argument, but that it be compatible with duality to a certain degree of approximation which will have to depend on the precision of the experiments used for establishing the 'fact' of duality. 
- \cite{PKF}, page 316-325
\end{quote}

In fact, conceptual revision is possible even about experimental facts which are taken to be exact and perfectly valid, when the concepts to be revised are strongly entrenched in everyday thinking, and when the new concepts to be introduced (which are in fact {\it invented}  to establish a new theoretical paradigm) are incompatible with the existing ones. The historic example is again taken from Galileo's story:

\begin{quote}
[...] 
How else could it have been possible [...] to replace the Aristotelian physics and the Aristotelian cosmology by the new physics of Galileo and Newton? The only conceptual apparatus then available was the Aristotelian theory of change with its opposition of actual and potential properties, form and matter, the four causes and the like. This conceptual apparatus was much more general and universal than the physical theories of today for it contained a general theory of change, spatiotemporal and otherwise. It also seems to be closer to everyday thinking and was therefore more firmly entrenched than any succeeding physical theory, classical physics included. Within this tremendously involved conceptual scheme Galileo's (or rather Descartes') law of inertia does not make sense. Should, then, Galileo have tried to get on with the Aristotelian concepts as well as possible because these concepts were the only ones in actual use and as 'there is no use discussing what could be done if we were other [i.e. more ingenious] beings that we are?' By no means! What was needed was not  improvement, or delimitation of the Aristotelian concepts  to 'make room for new physical laws'; what was needed was an entirely new theory. Now at the time of Galileo human beings were apparently able to do this extraordinary thing and become 'beings different from what they were before' (and one should again realize that the conceptual change that was implied was much more radical than the conceptual change necessitated by the appearance of the quantum of action). Are there (apart from pessimism with respect to the abilities of contemporary physicists) any reasons to assume that what was possible in the sixteenth and seventeenth centuries will be impossible in the twentieth century? - \cite{PKF}, page 316-325
\end{quote}

This conclusion about quantum theory becomes yet another instance of a more general conclusion about scientific progress, in its empirical, theoretical and philosophical components:
\begin{quote}
This result is exactly as it should be. Any restrictive demand with respect to the form and the properties of future theories can be justified only if an assertion is made to the effect that certain parts of the knowledge we possess are absolute and irrevocable. Dogmatism, however, should be alien to the spirit of scientific research, and this quite irrespective of whether it is now grounded upon 'experience' or upon a different and more \lq aprioristic' kind of argument. - \cite{PKF}, page 316-325
\end{quote}  

Having established that the distinction between theories and (empirical) facts is not an absolute one and that the observational language is a theoretical one, PKF argues that the choice between realism and instrumentalism amounts to a strategic (and heuristic) choice in our pursue of a scientific understanding of nature. It has to do with which concepts (whether those grounding the (more) theoretical language or those grounding the (more) observational one) we decide to rely on when trying to improve our scientific understanding of the world (see also \cite{Kuby2018}). As with any strategic decision, we should always remain flexible and constantly adjust our strategies as we move forward and question the assumptions at the basis of them. We should, in other words, be \lq rationally tightrope-walking\rq.

In other words, preferring realism over instrumentalism, or vice versa, is not something that can be decided on philosophical grounds (especially philosophical arguments regarding theory comparison, which are hindered by incommensurability), nor on empirical grounds, but by its results in facilitating progress, on a case-by-case and tentative basis, even though the concrete implementation of a new general perspective can be difficult. And the fact that one cannot decide a priori in favor of one option over the other does not make the choice an arbitrary one either, because the consequences for scientific progress of taking one option or the other can be judged. One could say that heuristic value is a more important criterion than epistemological or metaphysical basis (on this point, see \cite{Nickles1987}). In PKF's words:

\begin{quote}
The [second] objection is that different ideals of knowledge cannot be realized equally easily. Taking the second objection first, we are prepared to admit that there may be psychological difficulties in inventing theories of a certain type, especially if metaphysical views are held which seem to recommend radically different theories. However, a stronger variant of the second objection is frequently used. According to this stronger variant, all our theoretical knowledge is (uniquely) determined by the facts and cannot be chosen at will. Against this objection we repeat that what is determined by the 'facts' is the acceptance (or rejection) of sentences which are already interpreted and which have been interpreted independently of the phenomenological character of what is observed. [...]
such a situation arises whenever a fairly general point of view was held long enough to influence our expectations, our language and thereby our perceptions, and when during that period no alternative picture was ever seriously considered. We may prolong such a situation either by explaining away adverse facts with the help of ad hoc hypotheses which are framed in terms of the points of view to be conserved; or by reducing more successful alternatives to 'instruments of prediction' which, being devoid of descriptive meaning, cannot clash (in the phenomenological sense) with any experience; or by devising a criterion of significance according to which such alternatives are meaningless.  [...] But we can also choose the opposite procedure, i.e. we can take refutations seriously and regard alternative theories, in spite of their unusual character, as descriptive of really existing things, properties, relations etc. In short, although the truth of a theory may not depend upon us, its form (and the form of our theoretical knowledge in general) can always be arranged so as to satisfy certain demands. 
- \cite{PKF}, page 34-36
\end{quote}

Once more, the only real enemy is dogmatism, in all its forms and critical (self-)assessment is mandatory and should be a constant element of our investigations, whether we initially decide to proceed within a  realist perspective or from an instrumentalist standpoint (again, a similar understanding is expressed in \cite{Shaw2020}. One has to exercise tightrope-walking rationality, avoiding both a-critical realism and hypocritical instrumentalism, and keep going.

Having accepted this fundamentally anti-dogmatic stance, does PKF have arguments, however tentative and purely strategic in nature, for one or the other option? He does, and can be seen already from the last part of the previous quote, in which it is suggested to take a realist stance on alternative theories, even when blatantly in contrast with established 'facts',  to check their heuristic value. It can be argued, he maintains, that realism tends to be less dogmatic than instrumentalism, since it does not take even empirical facts and observations for granted (for a discussion of this point, see \cite{Wray2015}). Indeed, it treats all elements of scientific theories as subject to critical analysis, and leaves more room for metaphysical speculation. In this sense, it is {\it always} to be preferred\footnote{It is interesting to notice that, starting from the same anti-dogmatic stance, and in fact based on a similar emphasis on the existence of alternative explanations for the same phenomena, one can also be led to prefer an instrumentalist rather than realist attitude. This is the case, for example, of K. Stanford's version of instrumentalism \cite{Stanford}.}. This is a strong conclusion.

Here are PKF's words\footnote{Incidentally, the following quote is a justification of creativity in theory construction unhinged by empirical constraints and of non-empirical theory assessment, and a reminder that empirical evidence remains the ultimate judge of of our theories.}:

\begin{quote}
As opposed to positivism, a realistic position does not admit any dogmatic and incorrigible statement into the field of knowledge. Hence, also, our knowledge of what is observed is not regarded as unalterable and this in spite of the fact that it may have a counterpart in the phenomena themselves. This means that at times interpretations will have to be considered which do not 'fit' the phenomena and which clash with what is immediately given. Interpretations of this kind could not possibly emerge from close attention to the 'facts'. It follows that we need a non-observational source for interpretations. Such a source is provided by (metaphysical) speculation which is thus shown to play an important role within realism. However, the results of such speculation must be made testable, and having been transformed in this way they must be interpreted as descriptive of general features of the world (otherwise we are thrown back upon the old account of what is observed). This procedure (a) allows us to draw a clear boundary line between objective states of affair and the states of the observer, though it admits that we may be mistaken with regard to the exact position of the boundary line; it (b) is empirical in the sense that no dogmatic statements are allowed to become elements of knowledge; it (c) is liable to encourage progress by admonishing us to adapt even our sensations to new ideas; and it (d) allows for the universal application of the argumentative function of our language, and not only for its application within a given frame which itself can only be either described, or expressed - \cite{PKF}, page 34-36
\end{quote}

The general conclusion can then be applied to the analysis of the orthodox Copenhagen view of QM (in its simplified instrumentalist characterization), with the historical comparison with the attitude taken by many toward the Copernican system:

\begin{quote}
[...] It is also true that the quantum theory is the first theory of importance which to some extent satisfies the programme of Berkeley and Mach (classical states of affairs replacing the 'perceptions' of the former and the 'elements' of the latter). But it must not be forgotten that there is a whole tradition which is connected with the philosophical position of realism and which went along completely different lines. In this tradition the facts of experience, whether or not they are now describable in terms of a universal theory (such as classical mechanics), are not regarded as unalterable building stones of knowledge; they are regarded as capable of analysis, of improvement, and it is even assumed that such an analysis and improvement is absolutely necessary. [examples: Galileo and Copernico, atomic theory] [...] This tradition proceeds from the very reasonable assumption that our ideas as well as our experiences may be erroneous and that the latter gives at most an approximate account of what is going on in reality. Bohm's own point of view is closely connected with this tradition. 
 - \cite{PKF}, page 224-225
\end{quote}

Here we see PKF's preference for the realist view associated with Bohmian mechanics (quoting directly from Bohm), stressing how the latter  had to incorporate the lessons of QM and modify the concepts of classical physics to account for them, even while using the same concepts as the basis for a realist position. The new concepts are adopted {\it because a realist position is maintained} and this position forces us to question the immediate results of observations, and the concepts initially used to interpret them. An instrumentalist (or "positivist") attitude, PKF argues, makes this critical analysis of observational concepts more difficult, and ends up with \lq\lq a dogmatic restriction of the possible forms of future experience" (in Bohm's own words, as quoted in \cite{PKF}, page 224-225). 

\

So, realism is always to be preferred to instrumentalism, as a more fruitful (because less dogmatic) attitude for heuristic reasons, rather than a priori philosophical arguments.  It leads to questioning more our assumptions, it includes the inputs from experience among aspects of our scientific understanding of the world that require critical scrutiny, and it is supported by the realization that also our observations are loaded with theoretical elements. Let us analyse this strong conclusion further. 

\

First, for PKF, choosing between instrumentalism and realism means choosing a philosophical interpretation (of the relationship between theory and reality) that takes seriously the actual practice of science, i.e. what scientist do when they apply the theory in concrete experimental situations. This is a crucial aspect of most of PKF's production, in fact (in the context of PKF work on QM it has been noticed and analysed, for example, in \cite{Kuby}). In this way, we argue, instrumentalism and realism, and the choice between them, become (parts of) methodological rather than metaphysical positions.  This happens naturally, for several reasons. Actual scientific practice is, for PKF, the main basis as well as testing ground for any methodological position, rather than any abstract principle of rationality. A good methodology is one that takes on board all the elements that help, in a given context, advancing science. This is at the descriptive level but alo at the normative level. Thus, if scientist do make progress by adopting a given interpretation, this is a main argument in favor of that interpretation. Also, while the specific interpretative framework one adopts may not directly influence the laboratory practices one uses (in this sense it may sound misplaced to see the adoption of the framework as part of a methodology), sometimes it does, focusing attention on one or another phenomenon or specific techniques. A given interpretation may also make it easier to consider new theoretical hypotheses, to modify the same theoretical framework or appreciate the value of a different one. In this sense, flexibility at the interpretative level (even leaving laboratory practices untouched) can facilitate the application of key methodological principles and associated practical steps, like the principle of proliferation, so dear to PKF. PKF recognised  at several points that empirical assessment is only part of theory assessment (although the most solid and decisive one) and laboratory practices only part of theory development, while conceptual shifts and interpretative frameworks can play substantial role in both, so considerations about them are inevitably about methodology as well (see \cite{Shaw2020-2} for a related discussion).   
The key aspect to be appreciated in PKF's analysis is that realism is understood as a {\it methodological} position, and is supported as such, rather than a metaphysical position. It takes seriously, at the methodological level, but does not commit to, at the metaphysical level, the representational aspects of physical theories and the role they play in scientific practice. 

Second, we suggest that, by turning realism into a methodological position, PKF is anticipating R. Giere's constructive realism (emphasizing the nature of scientific theories as human constructions as well as recognizing the absence of sharp distinction between theoretical and observational elements in them) \cite{GiereConstructiveRealism}, but even more his proposed strategy of promoting any metaphysical presupposition into a methodological maxim, and then judging the metaphysical presupposition on the basis of its fruitfulness once translated into a methodological prescription \cite{GierePerspectivalPluralism}. This strategy, which appears to be entirely analogous to what PKF is proposing for realism in general, is a key element in Giere's formulation of perspectival realism \cite{GierePerspectivalPluralism} and it is also adopted by Giere in advocating for his more general programme of a "naturalised philosophy of science", where it is applied to the very notion of naturalism \cite{GierePhilScienceNaturalised}. The resonances between PKF's philosophy and Giere's have been also pointed out by Brown in \cite{Brown}, as well as by Giere himself \cite{GiereFeyerabend}, although mostly referring to PKF's later work. We point out that such resonances can be found in his earlier production. This \lq perspectival\rq ~aspect of realism\footnote{On the perspectival aspects and PKF's general stance on Realism see also Chang in \cite{modern2}}, apart from being a fertile and growing research direction in general philosophy of science \cite{massimi}, it has been applied by several authors to the interpretation of QM, mostly in relation to neo-Copenhagen perspectives (see e.g. \cite{Bitbol, Dieks}).  

Third, a {\it realist methodology}, i.e. the adoption of realism has some further important methodological implications. In essence, it seems to amount to "taking scientific theories seriously" for what they seem to say about the world, and explore their consequences fully, even when counterintuitive and conceptually weird or even in contradiction with observations while being ready to reinterpret and modify the observational concepts if it seems useful to do so. When considering more than one theory at once, on the other hand, there is more to gain from a realist methodology. The natural consequence of methodological realism is that different theories (about the same physical domain), taken seriously and pushed to their extreme consequences, will usually conflict radically in their underlying (suggested) ontology. Insisting on a realist attitude, instead of retreating to an instrumentalist one (i.e. noticing the radical conflict in their core ingredients but being content with their agreement with observations, which of course only forces to accept a very limited part of them), implies adopting a further key methodological principle, at the core of PKF's philosophy of science: proliferation (see the introduction to part 2 of \cite{PKF}, chapters 3 and 4 in \cite{PKF3}, and \cite{Tsou}. We need to be ready to pursue equally several competing theories, even when in radical conflict with each other and based on contradicting principles, coping with the contradiction and aiming for scientific progress and better understanding, rather than unity and agreement at the expenses of the potential progress, until the greater fruitfulness of one of them is established (and, PKF would argue, even afterwards). This proliferation of theories and research programmes, even contradicting one another at their core, is beneficial. There are many arguments for this conclusion, which was forcefully pushed by PKF, but has an earlier origin (dating back, as PKF himself notices, to J. Stuart Mill \cite{Lloyd1997}) and an independent academic lineage. Most of these arguments focus on the empirical assessment of theories and the discovery of observational anomalies, being often the direct result of the existence of alternatives to the given theory, and being at times invisible without such alternatives (see \cite{Hoyningen-Huene2000-HOYPKF-2}). The importance of proliferation, however, can also be argued for in the context of non-empirical theory assessment \cite{NoAlternativeProliferation}, although of course on a different basis. None of these argument, in either context, is conclusive and the principle of proliferation remains controversial. See \cite{Bschir2015-BSCFAP} for a review and more references.  

\

At this point, once the arena of the debate between realism and instrumentalism has become that of scientific methodology, one could and probably should use the same arguments in favor of instrumentalism, rather than realism, as PKF does. In other words, PKF dismissal of "instrumentalism" is correctly directed at dogmatic, positivist, excessively restrictive versions of it, but risks removing a useful methodological tool (contrary to his general stated goals). This would also follow naturally from PKF's own methodological anarchism, in fact.

Indeed, modern work in the foundations of QM has benefited greatly from the methodological advantages of "instrumentalism", in particular in its closely related version of "operationalism". The latter can be understood as the focus on the elements of QM that directly correspond to or follow from what agents (experimentalists or observers) actually do in the lab (intended in an idealized sense) when they manipulate (quantum) physical systems, somehow understanding the theory as a way to encode efficiently the results of such manipulations and its core principles as to be identified as stemming from them. This way of proceeding is most apparent in all the work on quantum foundations that proceeds in parallel with advances in quantum information and quantum computation, which provide at once a testing ground of foundational ideas and a source of inspiration for clarifying the foundations of QM. It can be seen in a lot of recent reconstructions of QM from operational principles, e.g. \cite{Hardy, Mueller} among many others, as well as in the development of a number of information-theoretic QM interpretations, e.g. \cite{RQM, BruknerZeilinger, QBismInfo} to mention just a few, which indeed are best understood as "neo-Copenhagen", inspired and further developed from the Bohr's tradition \cite{QBismBohr}. We will say more about these modern developments in the next section. For now, it should be clear that "instrumentalism" and "operationalism" can be very fruitful methodological positions. Indeed, this is also how most of the scientists involved in the cited developments use them, without necessarily endorsing them as general philosophical positions but as a convenient conceptual framework to force ourselves to focus on observable, testable, "experienceable" aspects of scientific theories, and thus bring to the forefront some of their essential elements. 
 
So, was PKF himself being too dogmatic in his arguments for realism, and in his conclusion that \lq\lq realism is always to be preferred to instrumentalism"? Yes, he was, and in fact, he noticed too. 

In editing his collected papers from this period, i.e. \cite{PKF}, he came to the same conclusion. Here we see a clearest example of tightrope-walking rationality in action, with PKF realizing that he is falling on one side of the rope and quickly rebalancing.
In the introduction to the volume, he states:

\begin{quote}
Chs. 2 and 11 [DO: containing some of the PKF quotes we reported above] are therefore somewhat misleading. Producing philosophical arguments for a point of view whose applicability has to be decided by concrete scientific research, they suggest that scientific realism is the only reasonable position to take, come what may, and inject a dogmatic element into scientific discussions (this dogmatism is responsible for the less than satisfactory nature of discussions about the foundations of the quantum theory). Of course, philosophical arguments should not be avoided; but they have to pass the test of scientific practice. They are welcome if they help the practice; they must be withdrawn if they hinder it, or deflect it in undesirable directions. - \cite{PKF}, page 15-16
\end{quote}

In this context, PKF again contrasts the philosophers' attitude to the physicists' more pragmatic one, praising the latter, stressing again that  physics arguments trump philosophical ones, however convincing. We thus come back to the discussion and conclusions of the previous section, on the general relation between philosophy (of physics) and physics. It is on this basis, as we have already noticed, that PKF came to gradually appreciate more and more Bohr's philosophical and physical position, albeit an "instrumentalist" one. And it is in the context of a reassessment of Bohr's position that PKF offers further interesting insights on the foundations of QM, of a more specific kind (i.e. not having to do with the general discussion on realism vs instrumentalism, and the nature of physical theories). Some of them are by now standard understanding, accepted points in the domain of quantum foundations, and probably not because of a direct influence of PKF's writings (whose impact on the philosophy of QM appears to have been rather limited). However, we find extremely interesting to notice their presence already in his (early) work. We now turn to a discussion of these insights. 
 
 \section{Specific insights into the foundations of QM}
The insights about QM from PKF that we find particularly interesting do not directly concern specific advances in the formalism or its use, even though in his writings one can find a number of careful analyses of QM results and some very useful comments about the quantum-classical transition, the relation between classical and quantum mechanics, and even some elements that seem to anticipate modern decoherence. 

The most interesting insights are found in his exposition of Bohr's views. Going beyond any simplification often applied to \lq\lq the Copenhagen interpretation" that reduces it to the most naive form of instrumentalist attitude, if not to the brutal \lq\lq shut up and calculate" mantra of many QM textbooks, PKF identifies and discusses at length some key components of Bohr's position and appreciates their conceptual depth.
 
 \
 
The main one is a clear understanding of quantum states as {\it relational} in nature, rather than individually associated to physical systems as intrinsic (encoding of) properties. In QM, the states associated to a physical system should always be understood in relation to a context, which in turn can be seen as corresponding to an observer or, more precisely, a measurement context. Because of this fundamentally relational nature, it becomes impossible to separate observer and observed system, and go back to an understanding of 'phenomena' as revealed by observation but independent of the act of observation, or, more generally, the interaction between two (observer + observed) systems. This relational nature is not to be misunderstood (as often done in popular accounts of QM) as meaning that the interaction with the observer or the measurement apparatus, i.e. its embodiment, affects or modifies or disturbs the properties of the observed system, to which one associates quantum states. This picture assumes that the two systems in interaction and their properties can be characterized separately and their properties separately attributed. The message of QM, according to Bohr (mediated by PKF), is much stronger. Phenomena cannot be {\it separated} in individual components pertaining to the observed system and those pertaining to the observer (or measurement context). Using PKF's words, in a section titled explicitly \lq\lq The relational character of quantum states":

\begin{quote}
[Complementarity] takes position, momentum, and all the other dynamical variables out of the individual physical system and attributes them to the experimental arrangement. The only properties which remain in the object and which can be attributed to it independently of the situation are mass, charge, baryon-number and similar 'non-dynamical' characteristics. We shall say that complementarity asserts the relational character not only of probability, but of all dynamical magnitudes. One of the consequences of this procedure is that 'phenomena' cannot be 'subdivided' ([2], 50), which just means that dynamical magnitudes cannot be separated from the conditions of their application: 'this crucial point. . . implies the impossibility of any sharp separation between the behaviour of atomic objects and the interactions with the measuring instrument which serve to define the conditions under which the phenomena appear' ([12], 39n°). Another consequence is that there arises an 'ambiguity in ascribing customary physical attributes to atomic objects' ([12], 51) - for these attributes no longer apply to the object per se but to the whole experimental arrangement, different assertions being made in different contexts. 
- \cite{PKF}, page 260-261
\end{quote}

And to clarify the inseparable nature of phenomena as well as their "objective, if relative" nature, described in analogy with the relativity of mechanical quantities in terms of reference frames in relativistic physics:

\begin{quote}
[...] Bohr has emphasized the need to refer the statistical results of the quantum theory to experimental conditions and to explain some of their changes not by a causal influence, but by a change of these conditions. It is in this spirit that he criticizes phrases such as 'disturbing the phenomena by observation', etc., which suggest that the reduction of the wave packet is due to an interaction between system and experimental conditions, both being regarded as separate entities bound together by a strong force." (In the theory of relativity which also ascribes properties to the total situation rather than to the individual system such a 'mechanistic' interpretation is provided by the point of view of Lorentz. [...] ) He quite explicitly reminds us that one must distinguish between 'a mechanical disturbance of the system under investigation' and 'an influence on the very conditions which define the possible types of [statistical] predictions regarding the future behaviour of the system'; and he compares this latter influence with 'the dependence on the reference system, in relativity theory, of all readings of scales and clocks'. \cite{PKF}, page 250-251
\end{quote}

The latter, in particular, is a very interesting parallel, drawn by Bohr and emphasized by PKF, since it is a clear inspiration for more recent developments in the interpretation of QM, in the same tradition. For example, it is advocated  to motivate Relational Quantum Mechanics \cite{RQM}, and similar perspectives \cite{healey} combining the relational, context-dependent nature of quantum states, with an objective understanding of this dependence, trying to avoid subjectivist aspects or a role for higher-level agency in quantum mechanics, indicating a totally embodied, naturalised notion of observer. We will comment further on this point, below.

\

Now, it is clear that, as we discussed in the previous section, in Bohr we find a clear emphasis on the classical elements that we use to characterize the measurement context and thus the "observing system" in ordinary applications of QM. It is similarly clear that this emphasis brings Bohr's position closer to instrumentalism. However, it is also clear that it is far from any strong or naive instrumentalism and that, while referring heavily to the results of measurements and observations, it does not take them for granted as primitive notions not requiring a careful analysis, so much so that Bohr's analysis leads indeed to a drastic revision, in QM, of what we mean by \lq phenomena\rq. 

And one could even say that this blurred distinction between observer and observed is a physical counterpart of the epistemological point made by PKF of a blurred distinction between theoretical and observational language in his criticism of the double layer model on which the debate on realism vs instrumentalism is based (see the previous section).

This blurred distinction, in fact, does not save any naive realist position either, in the interpretation of QM put forward by Bohr (as understood by PKF). In this interpretation, what Bohr is doing is proposing a drastic revision of \lq\lq classical realism", where the label "classical" refers both to the picture of the world inherited by classical Newtonian mechanics and to traditional, strong forms of philosophical or scientific realism.
 It is not instrumentalism, but rather it should be understood as a starting point for developing a new form of realism. 
 
 PKF gives a very clear account of this key aspect of Bohr's views, and points to a number of important implications of it. The same aspect and some of its implications had been understood earlier by other scholars scrutinizing Bohr's views, notably G. Hermann \cite{hermann, hermannBook} who offered an insightful analysis of Bohr's complementarity from a Kantian perspective. These Kantian elements have been recently expanded upon also by Cuffaro \cite{Cuffaro} and, specifically concerning the relative nature of observations or, as discussed here, of quantum states, by Crull \cite{Crull}. 
 
 Before we connect this insight by PKF with modern developments in the foundations of QM, let us  expand on a few more interesting aspects of this relational views of quantum states.  

One relevant aspect is the following. It is thanks to this somewhat relaxed version of realism, following from a contextual, relational interpretation of quantum states, that Bohr's view solves the EPR paradox without renouncing locality nor the completeness of QM. This transpires in full form when studying the Einstein-Bohr debate (e.g. starting from the Solvay conference \cite{Einstein-Bohr-Solvay}), but it hinges on this specific departure from classical realism. In fact, this provides the key element for a general template for a solution of the EPR paradox or the interpretative conundrum offered by Bell's theorem, which will then be followed and further articulated (in different ways) by modern QM interpretation in the "neo-Copenhagen" tradition.

In PKF's words:

\begin{quote}
[..] Einstein, Podolsky and Rosen assume without argument that what we determine when all disturbance has been eliminated is a property of the system examined that belongs to it irrespective of the conditions of the experiment (this is the content of their famous 'criterion of reality'). As opposed to this Bohr maintains, in accordance with his hypothesis, that quantum-mechanical state descriptions are relations between systems and experimental arrangement and are therefore dependent on the latter. It is easily seen how this assumption deals with EPR. For while a property cannot be changed except by direct interference with the system that possesses the property, a relation can be changed without such interference [...] Hence, lack of physical interference excludes change of state only if it has already been established that positions and momenta and other magnitudes are properties of systems, belonging to them under all circumstances. But just this assumption had to be rejected and had to be replaced by Bohr's hypothesis; and it had to be rejected because of difficulties which are quite independent of the case of EPR. Bohr's solution of this case is therefore not only not ad hoc, but it adds further support to the idea of complementarity. - \cite{PKF}, page 292-293 
\end{quote}
 
 and
 
 \begin{quote}
 [...]  Now according to Bohr, a quantum-mechanical state is a relation between (microscopic) systems and (macroscopic) devices. Also a system does not possess any properties over and above those contained in its state description (this is the completeness assumption). This being the case it is not possible, even conceptually, to speak of an interaction between the measuring instrument and the system investigated. The logical error committed by such a manner of speaking would be similar to the error committed by a person who wanted to explain changes of velocity of an object created by the transition to a different reference system as the result of an interaction between the object and the reference system. - \cite{PKF}, page 308-310
\end{quote}

Next, the relational interpretation of quantum states, as we have said, injects a radical element of context-dependence in any realist perspective on physical theories\footnote{We will discuss further this weakening of realism in the rest of this section. Here we notice that this is a central aspect, both profound and problematic, of several modern interpretations of quantum mechanics in various ways linked to Bohr's and maintaining the relational character of quantum states. This is well recognized and subject of an extensive debate, at physical and philosophical levels in the literature on quantum foundations. For one example among many, focusing on one specific interpretation, see \cite{Brown2009-BRORQM}.}. This is a strong and rather obvious link between Bohr's philosophy, and PKF's reading of it, and modern developments in philosophy of science like structural realism \cite{StructuralRealism}, with its emphasis on relations as the ontological substratum of the world, but even more like the already cited perspectival realism of Giere \cite{GiereConstructiveRealism, GierePerspectivalPluralism} and, more recently, Massimi \cite{massimi}. The connections with the latter are probably stronger, also due to the lesser relevance of metaphysical considerations both in Bohr and modern perspectival realism (as already mentioned in the discussion of Giere's methodological realism).

It is also important to emphasize that, like in these modern versions of scientific realism, the relational nature of quantum states does not lead to any radical subjectivism or to the insertion of necessarily mind-related aspects in the description of nature, in Bohr's perspective. While the "observer system" necessarily enters the characterization of quantum states and thus quantum phenomena always refer, at least implicitly, to a subject or agent, both these notions are entirely naturalised and objectified, and embodied in a measurement apparatus (described in classical terms). In particular, there is no role to play for consciousness or higher-level agency, or for the personal experience of subjects. Even the notions of "knowledge" and "information" acquired by the observer in her interaction with the observed physical system (so relevant for some of the modern neo-Copenhagen interpretations) are not relevant in Bohr's view. We are dealing, we could say, with a "fully objectified subject"\footnote{Here one can see maybe some traces of Whitehead's thinking, already linked to the conceptual challenges of quantum mechanics in \cite{Epperson2004-EPPQMA}. Along the same interpretation of PKF's thinking, see also \cite{Farell2001-FARFMP}.}. 
This should also be clear in light of the analogy, used by Bohr himself, and emphasized by PKF in the quote reported above, between the relative nature of quantum states, dependent on the measurement context, and the relative nature of mechanical quantities (e.g velocity) in relativistic physics, dependent on the reference frame. This is a parallel often made by proponents of Relational Quantum Mechanics, and one could speculate that PKF could have seen, in this risky attempt at a middle ground between subjectivism and objectivism, another instance of tightrope-walking at a metaphysical and epistemological level.

PKF is very careful in emphasizing this point, in defending Bohr's views from Popper's attack as excessively prone to subjectivism\footnote{It is also interesting to note how PKF emphasizes the Kantian aspects of Bohr's position and of the new relational understanding of quantum states and weaker realism, in line with what we also discussed above citing Hermann as well as recent scholars like Crull and Cuffaro.}:

\begin{quote}
Bohr is interested in the interaction between the subject and the object and he also emphasizes the similarity between physics and psychology in which latter science 'we are continually reminded of the difficulty of distinguishing between subject and object' ([8], 15 — original italics). But the 'subject' in physics is for him not the consciousness of the observer but 'the agency' used for observation, that is the material measuring instrument (including the body of the observer, and his sense organs); and the 'boundary' dis- appears, in physics, not from between the consciousness of the observer and 'the world'; it disappears from between 'the [atomic] phenomena [and] the [material] agencies of observation' ([8], 54). [...]  - \cite{PKF}, page 280-282
\end{quote}

So, the subject/observer in QM is fully embodied in a measurement context, but at the same time the object/observed necessarily looses its absolute, independent characterization. The absence of subjective (let alone mental/conscious) and epistemic elements, in this relational view, is repeatedly stressed by PKF:

\begin{quote}
[the relational character of the quantum-mechanical states] 
does not introduce any subjective element but concerns the physical situation only [...]. The hypothesis has certain similarities with Kantianism, assuming that nature (and the word is here used in almost the same sense as in Kant) depends on our categories and forms of perception. The difference is that the physical interactions involved in any act of cognition are taken into account and that consequences for epistemology are drawn from such physical considerations
[...]. - \cite{PKF}, page 280-282
\end{quote}

and again:

\begin{quote}
 Finally, it is evident that the 'observer' and his 'knowledge' nowhere enter the scene. All that is asserted is that there are objective conditions which do not allow for the application of certain magnitudes (and of the theories relating these magnitudes to each other) irrespective of whether these objective conditions are now used for improving our knowledge or not.
 to make this feature of the hypothesis as clear as possible I shall now illustrate it, as did Bohr ([9] and [88]), with the help of the very similar situation in relativity. [...] An object may be heavy in one frame, light in another, i.e. $M[0, R']$ may be $M[0, R"]$ and the value changes from one to the other depending on whether we consider the one or the other reference system. [...]   But again, it would be quite mistaken to try explaining the change in a 'mechanistic' fashion, as the result of an interaction between 0 and R (as is attempted in the electron theory of H. A. Lorentz); and we must again recall Bohr's warning not to confuse 'a mechanical disturbance of the system under consideration' with an 'influence on the very conditions' of the experiment. - \cite{PKF}, page 280-282
\end{quote}

All of the above suggests a novel form of relational objectivity.

It is also in this "relational objectivity" that one can see the connection between the relative nature of quantum states proposed by Bohr (that is, "complementarity") and the proposal made by Popper (and others) of an objective nature of probabilities (including quantum probabilities) in terms of propensities,  most clearly. 

Before we expand on this connection, let us stress that it is inevitable to inquire about the connection between how one interprets QM states and how one interprets probabilities, and a coherence between the two can even be demanded as a consistency condition of one's overall picture of the world. On the one hand, all that QM predicts are probabilities of the results of future measurements or, in a more "realist" ~take, the probabilities that the quantum systems (will be shown to) possess specific properties. This also impacts how we understand probabilities immediately enters how we interpret QM and vice versa, generally speaking. On the other hand, more specifically, quantum states can be directly traded for a list of probability assignments, both mathematically (one can reformulate the whole of QM replacing quantum states with probability assignments for the possible values of the various observables characterizing the system) and conceptually (e.g. holding that quantum states are just computational tools for real but probabilistic properties of systems, or that they simply encode observer's knowledge or information or belief about properties of systems, and it is this knowledge or information or belief that is real; that is, both in an objectivist and subjectivist understanding of probabilities). 
Thus, what interpretation of probabilities can be put in correspondence with Bohr's views and "orthodox" QM or with other QM interpretations? Vice versa, does a particular view of the nature of probabilities imply or favour a specific interpretation of QM? 

These questions are as difficult as they are interesting. Here we simply identify what PKF had to say about one specific instance of such relation, once more in a virtual debate with Popper, which apart from anything else shows how both Popper and PKF had already clear the strict relation between (the interpretation of) quantum states and probabilities. 

\

Popper \cite{Popper} had put forward his propensity-based interpretation of probabilities as a way to make them entirely objective properties of the world, thus removing subjective elements (e.g. in reference to knowledge of belief) from probabilistic theories like QM or classical statistical mechanics. It also had the goal, once this objective nature had been established and reference to "observers" or "agents" had been shown to be unnecessary, to support the view of QM as an incomplete description of reality, in the spirit of Einstein and against Bohr (and the whole Copenhagen "instrumentalist" perspective). Specifically, Popper maintained the classical realist picture in terms of particle trajectories and other observables taking definite values, even if they remain "hidden" in the "statistical" account provided by QM. The argument by Popper (as recounted by PKF) goes roughly as follows: many apparently surprising or even paradoxical features of QM are quickly recognised, once a propensity view of probabilities is adopted, as ordinary features of probability theory. This means that QM is simply a statistical theory. As a statistical theory, QM does not account for individual events but only for ensembles, while individual processes remain "hidden" from its description of the world, even if they remain in principle definite (as in classical mechanics); mistaking features of probability theory for features of a new mechanics of individual processes was the mistake of the founding fathers of QM, resulting from their having followed blindly a mistaken philosophy ("instrumentalism").

Upon analysis of Popper's arguments and Bohr's views, PKF concludes that the former has not offered, really, any cogent supporting argument for a classical realist picture hidden behind QM and that his criticism of Bohr's position on the basis of his propensity view of probabilities (the starting point of his argument) simply cannot be a criticism at all.

The main point that PKF makes, in fact, is that the distinguishing features of the propensity interpretation of probabilities not only do not contradict anything that Bohr has affirmed concerning quantum states or quantum systems, but, on the contrary, correspond exactly to the novel features identified by Bohr and made by him central to his view of QM. 
To put it simply, the distinguishing feature of propensities, i.e. the fact that they make (statistical) properties of systems dependent on (relative to) the conditions of observations, i.e. the measurement context, coincides with the relational nature of quantum states in Bohr's view, i.e. with complementarity, as discussed above:

\begin{quote}
The propensity interpretation [...] takes probabilities out of the individual physical system and attributes them to the experimental arrangement. Complementarity does the same - \cite{PKF}, page 260
\end{quote}

More in detail, it is exactly this feature of QM that Bohr takes as encoding the basic insight that "phenomena" are not something that can be separated into properties of individual systems independent of the context of observation. This aspect of the propensity view of probabilities, then, cannot in any way be turned against Bohr's QM, since it is already at its very core. And again, there is nothing "subjective" about it, as there is none in Popper's propensities, it is however a radical revision of our concept of physical phenomena, akin to that introduced by relativity, in the more limited domain of mechanical motion.

From the rest of the passage already quoted above:

\begin{quote}
as long as we consider probabilities a particular system cannot be separated from its surroundings, not even in thought. The probability of a die that has been shaken in a randomizer cannot therefore be explained as the result of an interaction between the die and the randomizer, or of a disturbance of the die by the randomizer. Any such account assumes that die and randomizer are separate systems which may affect each other in various ways. And while this is a perfectly adequate assumption in dynamics it must not be made in the case of probabilities which apply to the total experimental set-up. [...] The reader who is familiar with complementarity will at once recognize the great similarity with the view just described. Ever since 1935 [...] Bohr has emphasized the need to refer the statistical results of the quantum theory to experimental conditions and to explain some of their changes not by a causal influence, but by a change of these conditions. - \cite{PKF}, page 249-252 
\end{quote}

the conclusion being that Popper is repeating, in apparent opposition, what was already indicated by Bohr:
 
\begin{quote}
[...] The very same dependence of statistical results on experimental conditions is emphasized, about ten years later, by Popper who writes: '[...] the experiment as a whole determines a certain probability distribution'  
[...] 'The experiment as a whole' - that is precisely what Bohr means to express by his notion of a phenomenon which is supposed to 'refer to the observations obtained under specific circumstances including an account of the whole experimental arrangement' . [...] complementarity and the propensity interpretation coincide - as far as probabilities are concerned. In both cases probabilities are properties of 'blocks', be they now called 'experimental arrangements', or 'phenomena'. They are not properties, not even tendencies, of individual physical systems.  - \cite{PKF}, page 249-252
\end{quote}

Thus, PKF concludes, Popper's starting point, albeit correct, is definitely not enough to conclude that QM is just a statistical theory (only partially accounting for an underlying classical reality) and that its apparently novel, when not paradoxical, aspects are just features of probability theory. PKF, indeed, argues correctly that, even granted the probabilistic character of QM and its relational or contextual nature (in the sense of propensities or complementarity), none of the conceptual difficulties presented to us by QM are resolved: uncertainty, superposition and quantum fluctuations, not to mention entanglement; and none of the physical arguments and experimental results that directed the Copenhagen school away a classical realist position is made less relevant or explained differently by the propensity view. Propensities can be made part of QM, and this is indeed what Bohr did, but QM still requires a drastic rearrangement of our picture of the world, which is what the efforts of Bohr and his colleagues struggled to achieve.

Here is how PKF puts it:

\begin{quote}
It remains to say a few words about Popper's contention that 'the reduction of the wave packet [...] has nothing to do with the quantum theory: it is a trivial feature of probability theory'. The 'feature' which Popper has in mind is of course the dependence of probabilities on experimental conditions which entails that they change abruptly when the conditions change. Now the fact that the Born Interpretation is part of the quantum theory entails that this feature is part of the quantum theory also [...]. But while Popper, having before himself the ready-made result of the labours of the Copenhagen school, can regard the situation as a 'trivial feature of probability theory', Bohr has shown how difficult it is to have both the quantum theory and a propensity interpretation of its predictions, and how many modifications are needed  to achieve a combination of this kind 
 - \cite{PKF}, page 259-260
\end{quote}

On this point, PKF is correct, in our opinion. Funnily enough, in the same context, he is missing another side of Popper's position, which is important for modern interpretations of QM and for better appreciating Bohr's views. It seems to us, in fact, that in the quoted sentence, Popper had identified a key aspect of any interpretation of QM in which this is understood as being intrinsically about probabilities and in which quantum states are simply encoding of the same probabilities, devoid of an ontic status (they are not, that is, (functions of) real, intrinsic properties of the system they are associated to). In any such interpretation, the infamous \lq\lq measurement problem" is automatically solved or, better, dissolved without any need to modify or complement the QM formalism. In an epistemic interpretation of quantum states, in which they encode knowledge or information or belief about the properties of the system,  the collapse of the wave function associated to measurements, thus to the acquisition of new information about the system, is simply the updating of our knowledge and beliefs. The same is true in classical statistical mechanics, where the probability distribution over the classical phase space (or configuration space) also "collapses" to a new one upon measurement, i.e. it has to be substituted by a new one to reflect the newly acquired information. In a more objective perspective corresponding to a propensity interpretation of probabilities (as in Popper's), it corresponds to an objective change of the context in which the system is defined and on which the probabilities  depend, as we have seen. So, while PKF is right in stressing that the propensity view or a statistical interpretation of QM does not resolve all the conceptual puzzles it poses to our classical view of the world, and that the hard-won results of Bohr and co. need to be properly acknowledged.  But he is mistakingly downplaying this key insight by Popper about a natural resolution of at least one such puzzle offered by the propensity view\footnote{One could provocatively say that PKF, for once, has failed to be properly tightrope-walking between the severe critique of Popper and a proper appreciation of the insights Popper had to offer on the subject.}.

\

This avenue for a resolution (or rather, dissolution) of the measurement problem, in fact, is chosen by a large number of modern QM interpretations. It is chosen, for example, in the so-called epistemic approaches of "hidden variables" kind, formalized in the mathematical framework of "ontological models" \cite{Leifer}, already mentioned above. States of knowledge, in this context, naturally correspond to probability assignments for physical properties of the same system, normally understood in a Bayesian sense. While the first part is in line with Popper's desiderata, the Bayesian understanding of probabilities is not, and it could in principle be substituted with a propensity view.

Closer to the views of Bohr, the so-called neo-Copenhagen perspectives on QM dissolve the measurement problem by the same interpretative step. But, just like in Bohr, as correctly noted by PKF, this step does not imply instrumentalism, but rather points to a new form of realism, radically departing from the classical one. Such neo-Copenhagen perspectives on QM include (among others) the already mentioned Relational Quantum Mechanics \cite{RQM}, QBism \cite{QBismInfo}, the approaches of Brukner-Zeilinger \cite{BruknerZeilinger} and M\"{u}ller \cite{Mueller}, inspired by information theory, and Healey's pragmatism \cite{healey}. They differ in many aspects, but they can all be understood as taking onboard the key elements of Bohr's view that had been identified by PKF, while dropping the requirement that the observation context or the observer herself have to be characterized in classical terms, as Bohr insisted. We could argue that also this last point of departure from Bohr was hinted at by PKF, when criticizing the uncritical reliance on facts of experience and on the classical language used to express them, often present in naive instrumentalism. PKF's criticism would extend to the requirement that the classical characterization of results of observation, and of its context should necessarily resist to a generalization in QM, or that, in other words, the new \lq quantum realism\rq ~should find an inevitable boundary in this classical characterization of the observer or the observation. 

More precisely, freed from this limitation, the neo-Copenhagen interpretations of QM share: a) the relative nature of quantum states, giving to them an epistemic, as opposed to ontic, characterization\footnote{This could be in principle be understood in the "hidden variables" (or "statistical", in the wording of Popper), as implying a non-trivial link with ontic states encoding the actual reality of physical systems. This is not the case here. No such hidden reality is assumed and while quantum states are not themselves properties of systems, they do not need to be supplemented by anything.}\footnote{While this seems to apply to all \lq neo-Copenhagen\rq ~interpretations, it must be added that interpretations in this broad class differ significantly from one another. In some of them (Relational Quantum mechanics being an example) the objective, albeit relational or contextual, nature of quantum states is stressed, and the role of \lq knowledge\rq downplayed; the result is possibly an ontic understanding of quantum states, in which they are however not intrinsic properties of individual systems but extrinsic ones associated to pairs of systems. This is reflected also in the other aspects of the interpretations, e.g, the nature of probabilities or the nature of the \lq observing system\rq \cite{EpistemicQM}.};  and thus a relative nature of quantum \lq facts\rq ~or phenomena, their context-dependence; we could label this as pointing to an epistemology of perspectival objectivity, where the context-dependence or relational aspects do not imply subjectivity but indeed a new form of (admittedly weaker) objectivity; consequently, b) a view of reality as constantly \lq moulded\rq ~by the act of observation, of interaction between observer and observed, each embodied and naturalised, but fully defined only in terms of their relations and interactions; we could label this as a metaphysics of participatory realism. Where they differ is in the articulation of what physical and conceptual features should be attributed to the \lq  observer\rq  system in the QM context. They differ in the extent to which it can be embodied and naturalised and any subjective elements removed from it. Relatedly, they differ in the interpretation of quantum probabilities, again ranging from a most subjective bayesian one to an objective propensity-based one. For more details about the conceptual aspects of neo-Copenhagen interpretations, and their comparison, see \cite{EpistemicQM}.  

The above would require (and deserve) much more analysis to be developed properly. Here, we limit ourselves to pointing out how our earlier discussion of PKF's insights concerning realism and experience, in general, and in the QM context, allowed us to identify some seeds of what can be seen in these interpretations of QM, and offers some further elements to use in their conceptual analysis.

 \section{Concluding remarks}
 For a working theoretical physicist and a philosopher of physics, PKF is always an interesting read, whether one agrees or not with his arguments or conclusions. The main reason is that one perceives distinctively the goal of helping scientific progress, rather that promoting one or the other philosophical perspective as such; it sets his priorities, but certainly does not imply (as it is often the case for scientists reflecting on their own theories) a lack of conceptual rigour or philosophical depth. PKF's job, as a philosopher, was to put this rigour and depth at the service of scientific progress\footnote{This is true in the context of philosophy of science, while at a more general level, scientific progress itself should be ultimately at the service of human progress, including freedom, happiness and other values, according to PKF. This a conclusion most forcefully advocated in his later production, and one basis (another being his \lq relativism\rq , however subtle and controversial, in epistemological matters, for limiting the power of \lq science\rq ~and scientists in society \cite{Feyerabend1987-FEYFTR}.}. This is also the result of his general views about the relation between physics (science, more generally) and philosophy, as we have discussed above. A key element, that we stressed several times, is the constant worry to fall prey to any form of dogmatism, a defining feature of his \lq tightrope-walking rationality\rq and another one that is (or should be) natural to working scientists too. We hope to have made this apparent in our illustration of the many insights that PKF offered in his work on the foundations of QM, in turn mostly a case study of his general work on the realism vs instrumentalism debate, itself central in PKF's philosophy of science. Personally, we very much value these insights. 
 
 We recognize their limitations. They mostly concerned an analysis of Bohr's views (with related discussions of the views of Bohm and other early quantum physicists) and of early puzzles of QM. They are also, to some extent, outdated. Unfortunately, PKF stopped working on the foundations of QM exactly when this area of research started becoming more fertile as a scientific domain, and we do not know what PKF thought of Bell's theorems, or of Kochen-Specker theorem, to name some key early result in this area; and of course PKF could not comment on more recent ones, like Psi-ontology theorems, the interface between quantum foundations and quantum information, the developments of Bohmian mechanics and collapse models, or the more recent epistemic or neo-Copenhagen interpretations of QM. The same inevitable limitations concern PKF's analysis of realism, another topic which has seen a number of new philosophical directions spurring in more recent times, including also some that could be linked to some of PKF's insights that we discussed, e.g structural realism with its emphasis on relations, or perspectival realism. It would be unfair to focus much on these limitations. It would also be unfair, in our opinion, to emphasize the fact that PKF's insights did not concur to build any coherent, comprehensive view on QM or realism, or that he did not complement, so to speak, his insightful criticisms with more constructive work. True, but this sounds less surprising and less disappointing in light of our previous remarks on the intended role for philosophy in its interactions with physics, according to PKF's.
 
So, we take PKF's insights for what they are: valuable contributions to the progress of science, to our understanding of it, and in particular to the foundations of QM, coming from someone who truly cared about scientific progress and who was truly passionate about what science had unraveled (and is still unraveling), especially when it is as disconcerting and puzzling, radical and provocative (to any accepted physical wisdom) as the new quantum world.

\section*{Acknowledgements}
We acknowledge funding from DFG research grants OR432/3-1 and OR432/4-1, as well as from FQXi grant FQXi- RFP-IPW-1908. We thank A. Barzegar, S. Hartmann and E. Curiel for useful discussions on these topics. We thank E. Margoni for a careful reading of the first version of this paper, for many useful comments, and for directing us to Hermann's work, as well as Peter Morgan for further interesting comments. We thank the two anonymous reviewers and the editors for their many useful comments and criticisms and for pointing out several ways in which the first version of this work could be improved. We also express our gratitude to the editors of this special issue for the invitation to contribute to it and for the remarkable patience demonstrated in dealing with us. 
 
\bibliographystyle{jhep}
\bibliography{FeyerabendQM}

\end{document}